\documentclass[prc,superscriptaddress]{revtex4}% Physical Review C  
  
\usepackage{epsf}  
  
\begin{document}   
  
\title{Erratum:  Assessment of uncertainties in QRPA $0\nu\beta\beta$-decay  
    nuclear matrix elements [Nucl.\ Phys. A 766, 107 (2006)]}

%\date{\today}   
  
\author{V. A. Rodin}   
\email{vadim.rodin@uni-tuebingen.de}  
\affiliation{Institute f\"{u}r Theoretische Physik der Universit\"{a}t  
T\"{u}bingen, D-72076 T\"{u}bingen, Germany}  
\author{Amand Faessler}  
\email{amand.faessler@uni-tuebingen.de}  
\affiliation{Institute f\"{u}r Theoretische Physik der Universit\"{a}t  
T\"{u}bingen, D-72076 T\"{u}bingen, Germany}  
\author{F. \v Simkovic}  
\email{fedor.simkovic@fmph.uniba.sk}  
\altaffiliation{On  leave of absence from Department of Nuclear  
Physics, Comenius University, Mlynsk\'a dolina F1, SK--842 15  
Bratislava, Slovakia}   
\affiliation{Institute f\"{u}r Theoretische Physik der Universit\"{a}t  
T\"{u}bingen, D-72076 T\"{u}bingen, Germany}  
\author{Petr Vogel}  
\email{pxv@caltech.edu}  
\affiliation{Kellogg Radiation Laboratory 106-38, California Institute  
of Technology, Pasadena, CA 91125, USA}  

\maketitle

In a subsequent analysis a coding error was discovered in the treatment 
of the short range
correlations. Correcting the error resulted in  an increase  
of the neutrinoless double beta decay  matrix elements.
Here we provide the most relevant correct numerical results.
The corrected  Table
(replacing Table 1 of the paper) and the corrected figure 
(replacing Fig. 2 of the paper) are shown.

While the matrix elements are now larger, our basic claim that
the chosen way of adjusting the interaction strength makes the
matrix elements essentially independent on the size of the single particle
basis, on the parametrization of the G-matrix, whether QRPA or RQRPA is
used (although, as seen, QRPA results in $\sim$ 10\% larger matrix
elements than RQRPA), and whether $g_A$ is quenched or not remains true.

\begin{table}[htb]  
  \begin{center}  
%\squeezetable  
    \caption{Averaged $0\nu\beta\beta$ nuclear matrix elements  
 $\langle {M'}^{0\nu} \rangle$ and their variance $\sigma$  (in parentheses)  
evaluated in the RQRPA and QRPA. In column 6 the variance $\varepsilon_{exp.}$   
of the $0\nu\beta\beta$-decay matrix element due to uncertainties in the  
measured $2\nu\beta\beta$-decay half-live $T^{2\nu-exp}_{1/2}$   
is given. $M_{GT}^{exp}$ and $g_A$  
denote the $2\nu\beta\beta$-decay nuclear matrix element deduced from   
$T^{2\nu-exp}_{1/2}$  and axial-vector coupling constant, respectively.    
In column 7 the  $0\nu\beta\beta$ half-lives  
evaluated with the RQRPA average nuclear matrix element and for assumed  
$\langle m_{\beta\beta} \rangle$ = 50 meV are shown. For $^{136}$Xe there are four  
entries; the upper two use the upper limit of the $2\nu$ matrix element while  
the lower two use the ultimate limit, vanishing $2\nu$ matrix element.  
$^{150}$Nd is included for illustration. It is treated as a spherical nucleus;   
deformation will undoubtedly modify its $0\nu$ matrix element. }  
\label{tab:t12}  
\begin{tabular}{lcccccc}  
\hline\hline  
 Nuclear & $~g_A~~$ & $M_{GT}^{exp}$ & \multicolumn{2}{c} {$\langle {M'}^{0\nu} \rangle$} &  
 $~~\varepsilon_{exp.}~~$ &    
$T^{0\nu}_{1/2}$ ($\langle m_{\beta\beta} \rangle$ = 50 meV)  \\ \cline{4-5}  
transition & & [MeV$^{-1}$] & \hspace{0.3cm} RQRPA \hspace{0.3cm}   
& \hspace{0.3cm} QRPA \hspace{0.3cm}  & & [yrs] \\  
\hline   
$^{76}Ge\rightarrow {^{76}Se}$  
          & 1.25 & $0.15\pm 0.006$ & 3.92(0.12) &4.51(0.17) & $\pm0.05$ & $ 0.86_{-0.07}^{+0.08}~10^{27}$\\  
          & 1.00 & $0.23\pm 0.01$  & 3.46(0.13) &3.83(0.14) & $\pm0.06$ & $ 1.10_{-0.11}^{+0.13}~10^{27}$\\  
$^{82}Se\rightarrow {^{82}Kr}$     
          & 1.25 & $0.10\pm 0.009$ & 3.49(0.13) &4.02(0.15) & $\pm0.08$ & $ 2.44_{-0.26}^{+0.32}~10^{26}$\\  
          & 1.00 & $0.16\pm 0.008$ & 2.91(0.09) &3.29(0.12) & $\pm0.08$ & $ 3.50_{-0.38}^{+0.46}~10^{26}$\\  
$^{96}Zr\rightarrow {^{96}Mo}$     
          & 1.25 & $0.11^{+0.03}_{-0.06}$ & 1.20(0.14) &1.12(0.03) & $^{+0.12}_{-0.23}$ & $ 0.98_{-0.31}^{ +1.1}~10^{27}$\\  
          & 1.00 & $0.17^{+0.05}_{-0.1}$ & 1.12(0.11) &1.21(0.07) & $^{+0.12}_{-0.25}$ & $ 1.12_{-0.35}^{ +1.3}~10^{27}$\\ 
$^{100}Mo\rightarrow {^{100}Ru}$     
           & 1.25 & $0.22\pm 0.01$ &2.78(0.19) &3.34(0.19) & $\pm0.02$ & $ 2.37_{-0.32}^{+0.41}~10^{26}$\\  
           & 1.00 & $0.34\pm 0.015$ & 2.34(0.12) &2.71(0.14) & $\pm0.02$ & $ 3.33_{-0.39}^{+0.47}~10^{26}$\\  
$^{116}Cd\rightarrow {^{116}Sn}$     
           & 1.25 & $0.12\pm 0.006$ & 2.42(0.16) &2.74(0.19) & $\pm0.02$ & $ 2.86_{-0.39}^{+0.50}~10^{26}$\\  
           & 1.00 & $0.19\pm 0.009$ & 1.96(0.13) &2.18(0.16) & $\pm0.02$ & $ 4.39_{-0.61}^{+0.77}~10^{26}$\\  
$^{128}Te\rightarrow {^{128}Xe}$     
           & 1.25 & $0.034\pm 0.012$ & 3.23(0.12) &3.64(0.13) & $\pm0.09$ & $ 4.53_{-0.53}^{+0.64}~10^{27}$\\  
           & 1.00 & $0.053\pm 0.02$  & 2.54(0.08) &2.85(0.08) & $\pm0.10$ & $ 7.35_{-0.88}^{+1.1}~10^{27}$\\  
$^{130}Te\rightarrow {^{130}Xe}$  
           & 1.25 & $0.036^{+0.03}_{-0.009}$ & 2.95(0.12) &3.26(0.12) & $^{+0.26}_{-0.08}$ & $ 2.16_{-0.46}^{+0.33}~10^{26}$\\  
           & 1.00 & $0.056^{+0.05}_{-0.15}$  & 2.34(0.07) &2.59(0.06) & $^{+0.27}_{-0.08}$ & $ 3.42_{-0.83}^{+0.51}~10^{26}$ 
	   \\ 
$^{136}Xe\rightarrow {^{136}Ba}$ 
           & 1.25 & $0.030$ & 1.97(0.13) & 2.11(0.11) & & $4.55_{-0.56}^{+0.68}~10^{26}$ \\  
           & 1.00 & $0.045$ & 1.59 (0.09) & 1.70 (0.07) & & $6.38_{-0.91}^{+1.12}~10^{26}$ \\  
           & 1.25 & 0 & 1.67(0.13) & 1.78(0.11) & & $7.00_{-0.71}^{+0.84}~10^{26}$ \\  
           & 1.00 & 0 & 1.26 (0.09) & 1.35 (0.07) & & $1.11_{-0.14}^{+0.17}~10^{27}$ \\  
$^{150}Nd\rightarrow {^{150}Sm}$     
           & 1.25  & $0.07^{+0.009}_{-0.03}$ & 4.16(0.16) &4.74(0.20) & $^{+0.06}_{-0.19}$ & $ 2.23_{-0.21}^{+0.41}~10^{25}$\\  
           & 1.00  & $0.11^{+0.014}_{-0.05}$ & 3.30(0.16) &3.72(0.20) & $^{+0.06}_{-0.19}$ & $ 3.55_{-0.42}^{+0.87}~10^{25}$\\  
\hline\hline  
\end{tabular}  
  \end{center}  
\end{table}  
 
\newpage  
 
\begin{figure}[htb]   
  \begin{center}   
    \leavevmode   
    \epsfxsize=1.\textwidth   
    \epsffile{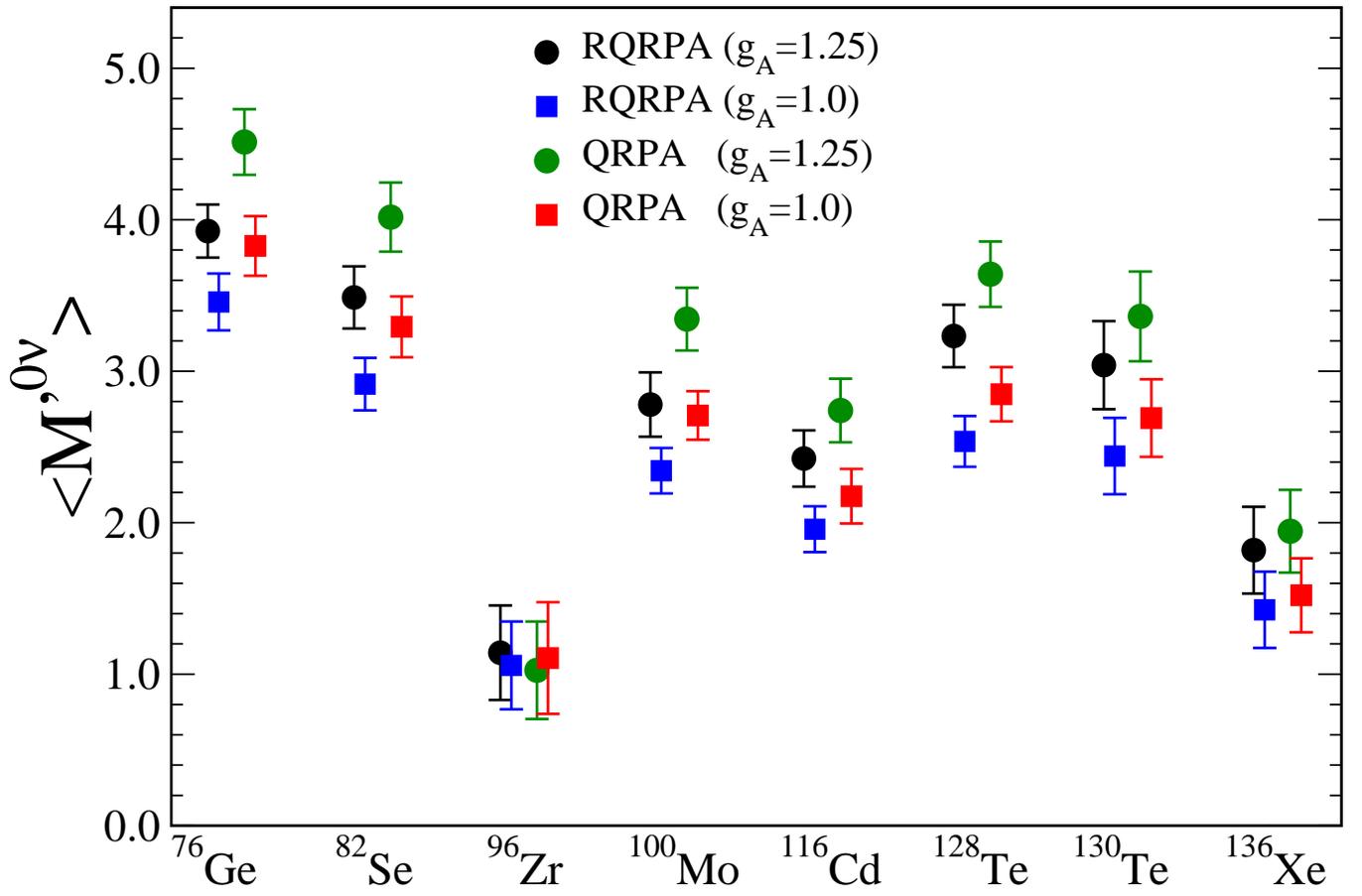}   
\vspace{0.5cm}   
    \caption{Average nuclear matrix elements $\langle {M'}^{0\nu} \rangle $   
and their variance (including the uncertainty coming from the experimental error in $M^{2\nu}$)   
for both methods and for all considered nuclei. For $^{136}$Xe the error bars encompass the whole interval related   
to the unknown rate of the $2\nu\beta\beta$ decay.}   
  \end{center}   
\end{figure}

\end{document}